\documentclass[twocolumn,showpacs,preprintnumbers,amsmath,amssymb,floatfix]{revtex4-1}

\usepackage{graphicx}
\usepackage{dcolumn}
\usepackage{bm}

\begin{document}

%

\let\a=\alpha      \let\b=\beta       \let\c=\chi        \let\d=\delta
\let\e=\varepsilon \let\f=\varphi     \let\g=\gamma      \let\h=\eta
\let\k=\kappa      \let\l=\lambda     \let\m=\mu
\let\o=\omega      \let\r=\varrho     \let\s=\sigma
\let\t=\tau        \let\th=\vartheta  \let\y=\upsilon    \let\x=\xi
\let\z=\zeta       \let\io=\iota      \let\vp=\varpi     \let\ro=\rho
\let\ph=\phi       \let\ep=\epsilon   \let\te=\theta
\let\n=\nu
\let\D=\Delta   \let\F=\Phi    \let\G=\Gamma  \let\L=\Lambda
\let\O=\Omega   \let\P=\Pi     \let\Ps=\Psi   \let\Si=\Sigma
\let\Th=\Theta  \let\X=\Xi     \let\Y=\Upsilon

%

%

\def\cA{{\cal A}}                \def\cB{{\cal B}}
\def\cC{{\cal C}}                \def\cD{{\cal D}}
\def\cE{{\cal E}}                \def\cF{{\cal F}}
\def\cG{{\cal G}}                \def\cH{{\cal H}}
\def\cI{{\cal I}}                \def\cJ{{\cal J}}
\def\cK{{\cal K}}                \def\cL{{\cal L}}
\def\cM{{\cal M}}                \def\cN{{\cal N}}
\def\cO{{\cal O}}                \def\cP{{\cal P}}
\def\cQ{{\cal Q}}                \def\cR{{\cal R}}
\def\cS{{\cal S}}                \def\cT{{\cal T}}
\def\cU{{\cal U}}                \def\cV{{\cal V}}
\def\cW{{\cal W}}                \def\cX{{\cal X}}
\def\cY{{\cal Y}}                \def\cZ{{\cal Z}}
%

\newcommand{\Ns}{N\hspace{-4.7mm}\not\hspace{2.7mm}}
\newcommand{\qs}{q\hspace{-3.7mm}\not\hspace{3.4mm}}
\newcommand{\ps}{p\hspace{-3.3mm}\not\hspace{1.2mm}}
\newcommand{\ks}{k\hspace{-3.3mm}\not\hspace{1.2mm}}
\newcommand{\des}{\partial\hspace{-4.mm}\not\hspace{2.5mm}}
\newcommand{\desco}{D\hspace{-4mm}\not\hspace{2mm}}



\title{Some Unfinished Thoughts on Strong Yukawa Couplings 
 }

\author{
Wei-Shu~Hou
 }
\affiliation{Department of Physics,
       National Taiwan University,
       Taipei, Taiwan 10617
 }

\date{\today}

\begin{abstract}
\noindent
Yukawa couplings of electroweak Goldstone bosons can be
inferred from experiment, but the existence of an elementary
Higgs boson is not yet an established fact.
If a sequential chiral quark generation does exist, it would
bring us now into the strong Yukawa coupling regime.
Guided by a Bethe--Salpeter equation approach,
we postulate that the leading collapse state, the
(heavy) isotriplet and color-singlet $\pi_1$ meson,
becomes the Goldstone boson $G$ itself.
Viewing it as a deeply bound state,
a gap equation is constructed.
This ``bootstrap" picture for electroweak symmetry breaking
relies on strong Yukawa coupling, without providing
any theory of the latter.
\end{abstract}

\maketitle

\section{Breaching Unitarity Bounds}

Particle physics has seen a great leap forward in 2011:
we celebrate the enormous success of the LHC,
running at 7 TeV.
But apprehension arose: No New Physics was seen, while
enormous parameter space was excluded for the Higgs boson,
the holy grail of LHC physics. To quote a sage~\cite{Peskin}:
``Our field seems to be approaching a definite point of reckoning.
But will it lead us to enlightenment, or to disillusionment and chaos?"

In a dramatic way, the strong hint~\cite{EPSHiggs} as of
late July for the (dreaded by most) $\sim 140$ GeV Higgs boson
diminished~\cite{LPHiggs} by end of August,
and dropped from view~\cite{Dec13Higgs} by December 13.
In its stead, there is a mild hint at 125 GeV, where,
though chastised by the (forced-by-LHC) retreat to above TeV scale,
the supersymmetry camp draws the  momentary warmth.

In this note we shun the 125 GeV Higgs possibility (we would know by
end of 2012 whether it shares the fate of the 144 GeV hint of July 2011),
but consider the other option: $m_H > 600$ GeV~\cite{Dec13Higgs}.
In fact, we would shun the whole idea of
the Higgs boson as an elementary particle, and
\emph{return to the basics of experiment-based knowledge}.
Simply put, we do not have any firm experimental knowledge
that the Higgs boson even exists.
Rather, it is the most economical theoretical construction
for electroweak symmetry breaking (EWSB),
at the cost of bringing in many theoretical problems, such as
quadratic divergences and the resulting hierarchy problem.

The physical bound of $m_H > 600$ GeV would imply one is
close to the onset of strong $W_LW_L$ scattering~\cite{LQT77}.
The approach to another ``unitarity bound" (UB) is also
imminent: strong $Q\bar Q$ (and $QQ$) scattering
at high energies for heavy chiral quark $Q$~\cite{CFH78}.
The CMS experiment has searched for the 4th generation
$t'$ and $b'$ quarks, and the stringent bounds~\cite{CMS4th},
at $\sim$ 500 GeV, are not far from the UB of
$\sim$ 500--600 GeV for $Q\bar Q$ scattering.

The 4th generation faces the difficulty of
unseen Higgs~\cite{Peskin,Higgs4},
given that it tends to enhance Higgs production via
gluon fusion by an order of magnitude. But taking
$m_H > 600$ GeV neutralizes this stigma~\cite{Denner}. Instead, could the
near UBV (UB violation) of strong $W_LW_L$ scattering
and strong $Q\bar Q$ (and $QQ$) scattering be correlated?
Could the \emph{strong Yukawa couplings} of a new sequential
heavy chiral quark generate~\cite{Holdom06} EWSB itself?
This is the theme we shall explore.

\section{From Gauge to Yukawa Couplings}

Given the curious absence of evidence for a Higgs boson,
let us recall the firm facts from experiment.

First, we know~\cite{PDG} that $q/l$
pointlike to $10^{-18}$ m , and governed by the
${\rm SU(3)_C} \otimes {\rm SU(2)_L} \otimes {\rm U(1)_Y}$
gauge dynamics. Chromodynamics would not be our concern,
but it is important to emphasize that, unlike the 1970s and early 1980s,
the ${\rm SU(2)_L} \otimes {\rm U(1)_Y}$ \emph{chiral} gauge dynamics
is now experimentally established.
We know that quarks and leptons come in left-handed weak doublets
and right-handed singlets, and for each given electric charge,
they carry different hypercharge $Y$.
%

Second, the weak bosons are found~\cite{PDG} to be
massive, $M_W = \frac{1}{2}gv$, where $g$ is the
measured ${\rm SU(2)_L}$ weak coupling, and $v^2=1/\sqrt{2}G_F$
the vacuum expectation value. Hence,
spontaneous breaking of ${\rm SU(2)_L} \otimes {\rm U(1)_Y}$
symmetry (SSB) is also experimentally established.

Third, all fermions are observed~\cite{PDG} to be
massive. These masses also indicate EWSB, since
they link left- and right-handed fermions of same electric charge,
but different ${\rm SU(2)_L}$ and ${\rm U(1)_Y}$ charges.
We shall not invoke the elementary Higgs boson for
mass generation, as it is not yet observed experimentally.

At this point we need to acknowledge the important theoretical
achievement of renormalizability~\cite{tHV72} of non-Abelian gauge theories,
which allowed theory-experiment correspondence down to
per mille level precision. Worthy of note is that the proof of
renormalizability is based on Ward identities,
and~\cite{'tHooft71} unaffected by SSB,
i.e. the underlying symmetry properties are not affected.
From this, we now argue~\cite{HouICHEP10} for the existence of Yukawa couplings
as an experimental fact.

\begin{figure}[t!]
\begin{center}
\vskip-2.1cm
 \includegraphics[width=90mm]{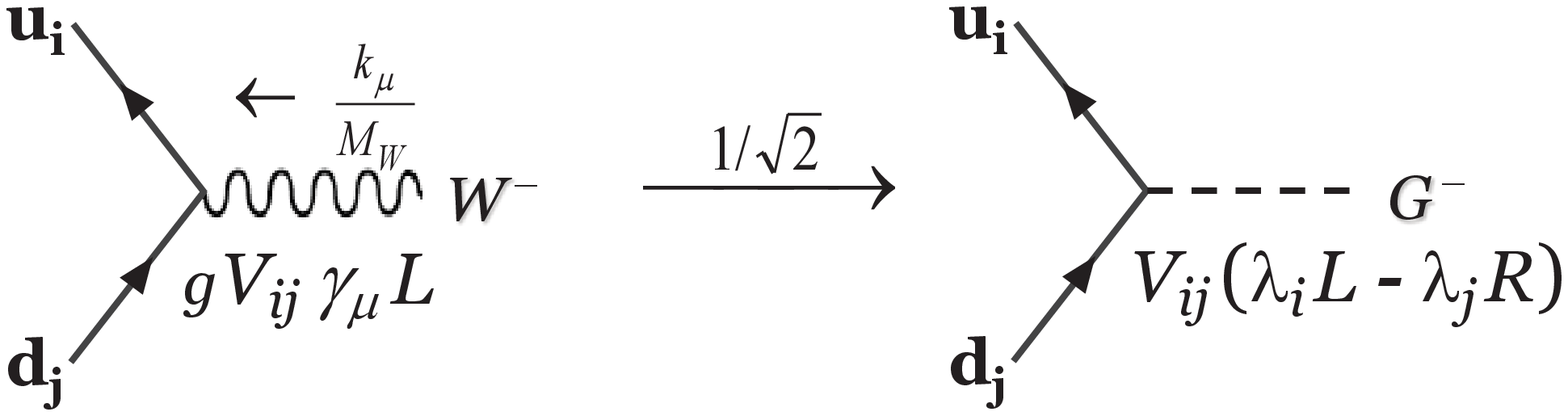}
\vskip-2.3cm
\caption{Deriving the Yukawa coupling,
 \emph{i.e.} Goldstone boson coupling to quarks,
 from purely left-handed gauge couplings.}
 \label{fig1}
\end{center}
\end{figure}

With proof of renormalizability, we choose the physical
unitary gauge, hence there are no would-be Goldstone bosons
(or unphysical scalars), only massive gauge bosons~\cite{KugoComment}.
The $\frac{k_\mu k_\nu}{M_W^2}$ part of a $W$ boson
propagator reflects longitudinal $W$ boson propagation
(which is the would-be Goldstone bosons that got ``eaten").
If we take a $\frac{k_\mu}{M_W}$ factor and contract with
a $d_j \to u_i$ charged current, as illustrated in Fig.~1,
simple manipulations give (dropping $V_{ij}$ for convenience),
\begin{eqnarray}
\frac{g}{\sqrt{2}} \frac{k{\hskip-0.18cm}/}{M_W} L
 &=& \frac{g}{\sqrt{2}} \frac{p{\hskip-0.18cm}/_i - p{\hskip-0.18cm}/_j}{M_W} L
  =  \frac{g}{\sqrt{2}} \frac{m_i L - m_j R}{M_W} \nonumber \\
 &=& \sqrt{2}\left(\frac{m_i}{v} L - \frac{m_j R}{v}\right) \nonumber \\
 &\equiv& \lambda_i L - \lambda_j R,
 \label{Gauge-Yuk}
\end{eqnarray}
where we have inferred
\begin{equation}
\lambda_Q \equiv \frac{\sqrt{2} m_Q}{v},
 \label{Yuk}
\end{equation}
as the effective $W_L$, or Goldstone boson $G$ coupling
to quarks, which is nothing but the familiar Yukawa coupling.
We have used the equation of motion in the second step
of Eq.~(\ref{Gauge-Yuk}), but this is justified since
we work in the broken phase of the real world, and
we know that all quarks are massive.

From Fig.~1 and Eqs.~(\ref{Gauge-Yuk}) and (\ref{Yuk}),
we see that from the experimentally established
left-handed gauge coupling, the Goldstone boson
couples via the usual Yukawa coupling.
The Goldstone bosons of EWSB pair with the
transverse gauge boson modes to constitute a
massive gauge boson, the Meissner effect,
but the important point is that we have not introduced
a physical Higgs boson in any step.
Unlike the Higgs boson, SSB of electroweak symmetry
is an experimentally established fact.
The Goldstone bosons couple with Yukawa couplings
proportional to fermion mass.

We have kept a factor $V_{ij}$ in Fig.~1.
Recall that the Kobayashi--Maskawa (KM) formalism~\cite{KM}
for quark mixing deals with massive quarks, or
equivalently the existence of Yukawa matrices, and
the argument remains exactly the same.
Vast amount of flavor and $CP$ violation (CPV) data
overwhelmingly supports~\cite{PDG} the 3 generation KM picture.
For example, the unique CPV phase with 3 quark generations
can so far explain all observed CPV phenomena.
These facts further attest to the existence of
Yukawa couplings from their dynamical effects,
but again do not provide any evidence for
the existence of the Higgs boson.

\section{Yukawa Bound States:  A Postulate}

Based on experimental facts and the renormalizability of
electroweak theory, we have ``derived" Yukawa couplings
from purely left-handed gauge couplings in the previous
section, without invoking an explicit Higgs sector,
at least not at the empirical, heuristic level.
We turn now to a more hypothetical situation:
Could there be more chiral generations?
Since we already have three, the possibility that
there exists a fourth generation of quarks
should not be dropped in a cavalier way.
Indeed, there has been some resurgent
interest~\cite{CMS4th, 4S4G09} recently,
and as argued in the introductory section,
one should press on when considering the ``heavy Higgs" scenario.
What we do know is that the 4th generation
$t'$ and $b'$ quarks should be suitably
degenerate to satisfy electroweak constraints
on the $S$ and $T$ variables~\cite{KPST07}.
A ``heavy isospin" is in accord with the
custodial SU(2) symmetry.

\begin{figure}[t!]
\begin{center}
\vskip-1.8cm
 \includegraphics[width=82mm]{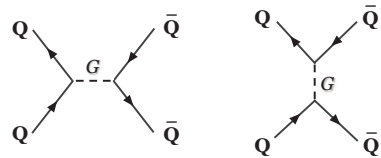}
\vskip-2.1cm
\caption{Scattering diagrams for $t$- and $s$- channel
 Goldstone boson $G$ exchange between $Q$ and $\bar Q$ quarks.
 Analogous diagrams can be drawn for $g$ (and $H$) exchange.}
 \label{fig1}
\end{center}
\end{figure}

With $m_{t'} \cong m_{b'} \equiv m_Q \gtrsim$
500 GeV~\cite{CMS4th}, their Yukawa couplings
are already 3 times stronger than the top quark,
hence stronger than all gauge couplings.
There has been two complementary studies of strong
Yukawa bound states. The first approach is along
traditional lines of relativistic expansion~\cite{Wise11}.
Ignoring all gauge couplings except QCD,
and taking the heavy isospin limit (hence
$Q$ represents 4th generation quark doublet, and
$G$ the triplet of Goldstone bosons), the
$t$- and $s$-channel Goldstone exchange diagrams
are depicted in Fig.~2, with corresponding diagrams
for $g$ as well as $H$ exchange
(Ref.~\cite{Wise11} did not put in $s$-channel gluon exchange).

The heavy $\bar QQ$ mesons form isosinglets
and isotriplets, and can be color singlet or octet.
We borrow the notation from hadrons and call these states
$\eta_1$, $\omega_1$, $\pi_1$, $\rho_1$ and
$\eta_8$, $\omega_8$, $\pi_8$, $\rho_8$, respectively.
Ref.~\cite{Wise11} used a variational approach,
with radius $a_0$ as parameter.
It was found that, for color singlet $\omega_1$ ($\rho_1$),
$a_0/a_{\rm QCD} \sim 1$ for $m_Q$ below 400 (540) GeV,
but above which $a_0$ suddenly precipitates towards tiny values.
For $\eta_1$ ($\pi_1$) the radius mildly decreases (increases)
from 1, with reverse trend for binding energy,
hence it remains QCD-bound.

To understand this, note that the
$t$-channel Goldstone exchange for $\eta_1$ is repulsive,
while the $s$-channel Goldstone exchange,
which contributes only to $\pi_1$, is also repulsive.
However, the sudden drop in $\omega_1$ and $\rho_1$ radii
is due to the trial wave function suddenly sensing
a lower energy at tiny radius due to $t$-channel
Goldstone exchange: the strong Yukawa coupling has
wrested control of binding from the Coulombic QCD potential.
QCD binding energy is only a couple of GeV, but the
sudden drop in radius leads to a sharp rise in
binding energy, giving rise to a kink.
The relativistic $v/c$ expansion fails just when it starts to get interesting.
For color octet states, QCD is repulsive, so $\eta_8$
does not bind.
In Ref.~\cite{Wise11}, the $\omega_8$ and $\pi_8$
states are degenerate, with sudden shrinking of radius
occurring around 530 GeV, but the $s$-channel QCD effect,
left out in Ref.~\cite{Wise11}, should push the $\omega_8$ upwards;
the $\rho_8$ state does not shrink until later.

Given that a relativistic expansion breaks down,
a truly relativistic approach is needed. Such a study,
based on a Bethe--Salpeter (BS) equation~\cite{Jain},
was pursued around the time of demise of the SSC.
The BS equation is a ladder sum of $t$- and $s$-channel diagrams
of Fig.~2, where the $\bar QQ$ pair forms a heavy meson
bound state. While the ladder sum of $t$-channel diagrams
are intuitive, a problem emerges for the $s$-channel,
which contributes only to $\pi_1$, $\omega_8$ and $\sigma_1$
(same quantum numbers as $G$, $g$ and $H$, respectively).
Rather than a triangle loop, the $s$-channel loop
appears like a self-energy hence potentially divergent,
while the momentum carried by the exchanged boson
is the bound state mass itself.
One could not formally turn the integral equation
into an eigenvalue problem.
This was resolved in Ref~\cite{Jain} by a subtraction
at fixed external momentum, which in effect eliminates
all $s$-channel diagrams.
Ref.~\cite{Jain} then solved the BS equation numerically
using several different approximations, which,
in addition to the approximate nature of the BS equation
itself, illustrates the uncertainties.
Still, unlike Ref.~\cite{Wise11}, the bound state masses
drop smoothly below $2m_Q$ as $m_Q$ increases,
showing no kink, which is an improvement.
However, a generic feature is \emph{collapse}: bound state
masses tend to drop sharply to zero at some high $m_Q$,
and would naively turn tachyonic.

Before we elaborate further about collapse, as well as
issues regarding subtracting out $s$-channel contributions,
we mention that a relative conservative study~\cite{EHY11} of
strong Yukawa bound state phenomenology was conducted
using the BS equation approach as a guide.
The mass range selected for study is $m_Q \in$ (500, 700) GeV,
where one would already have strong binding energy of order 100 GeV,
but still safely away from the region of collapse,
hence one could gainfully use the numerics of Ref.~\cite{Jain}.
Without solving the bound state problem in numerical detail,
the meson decay constant and other parameters were employed
to discuss LHC phenomenology in the near future.

Here, we do not pursue the phenomenology, but wish to
address more fundamental issues.
Although the \emph{de facto} $s$-channel subtraction made by
Ref.~\cite{Jain} appeared reasonable on formal grounds,
the contrast with the relativistic expansion is striking:
the Goldstone $G$ exchange in $s$-channel lead to
a specific repulsion~\cite{Wise11} for $\pi_1$ heavy mesons,
disallowing it to shrink suddenly like the
otherwise analogous $\omega_1$.
But after subtracting the $s$-channel, Ref.~\cite{Jain} finds
the \emph{$\pi_1$ as the most attractive channel} (MAC),
more so than $\omega_1$.
Together with the tendency towards collapse for large enough
$m_Q$ (equivalently $\lambda_Q$),
this means that the $\pi_1$ meson would be the
first to drop to zero and turn tachyonic.
That this occurs for the channel that experiences
repulsion when $2m_Q$ is far lower than collapse values
(\`{a} la $\omega_1$ which has no $s$-channel effect)
seems paradoxical.
Does this falsify the whole approach,
or else what light does this shed?
And how is it related to $s$-channel subtraction?

With experimental bounds~\cite{CMS4th} for 4th generation
quarks  entering the region of deep(er) binding,
we offer a self-consistent view that may seem a bit radical.
Clearly, around and below 500 GeV mass, or
$2m_Q \lesssim$ 1 TeV, there could still be some
repulsion due to $s$-channel $G$ exchange.
But since we did not introduce any elementary Higgs doublet,
the Goldstone boson $G$ should perhaps be viewed
as a $\bar QQ$ bound state. Hence, we
\begin{description}
 \item{Postulate}\;{\bf :} $\;\pi_1 \equiv G$, i.e.
 collapse is a precursor to dynamical EWSB,
 and the first mode to collapse becomes the
 Goldstone mode.
\end{description}

Although the full validity of the BS equation may
be questioned, it is known~\cite{Kugo78} that ``the appearance
of a tachyonic bound state leads to instability of
the vacuum", which is ``resolved by condensation
into the tachyonic mode".
Our Postulate removes the equation for $\pi_1$ self-consistently~\cite{omega1},
and provides some understanding of the $s$-channel subtraction:
a $\pi_1/G$ boson carrying $p^2 \sim (2m_Q)^2$ would
no longer be a bound Goldstone boson in $s$-channel.
Without an elementary Higgs boson,
there is no $\sigma_1$ channel subtraction,
while for heavy enough $m_Q$ (so $\pi_1$ has turned Goldstone)
one can treat QCD effects as a correction,
\emph{after} solving the $\omega_8$ bound state problem,
without need of subtracting $s$-channel $g$ exchange.
The self-consistent MAC behavior of the $\pi_1$ channel
seems like a reasonable outcome of the Goldstone
dynamics, as implied by the gauge dynamics.

It may now appear that EWSB is some kind of a
``bootstrap" from ``massive chiral quarks"
with large Yukawa coupling as seen in broken phase.

\section{A Gap Equation without Higgs}

Motivated by the previous heuristic discussion, we construct
a gap equation for the dynamical generation of heavy quark mass
without invoking the Higgs boson.

Connecting two of the
$Q$ or $\bar Q$ lines in Fig.~2, one gets
the self-energy for $Q$ by $G$ exchange, and readily
arrives at the gap equation as depicted in Fig.~3.
One treats both $Q$ and $G$ as massless at the diagrammatic level.
If quark mass $m_Q$, represented by the cross \textsf{X},
could be nonzero, then one has dynamical chiral symmetry breaking,
which is equivalent to EWSB!

\begin{figure}[t!]
\begin{center}
\vskip-2cm
 \includegraphics[width=90mm]{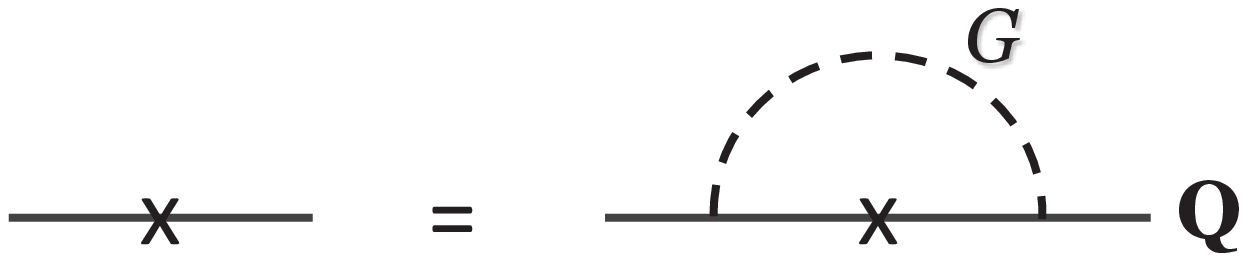}
\vskip-3cm
\caption{Gap equation for generating
heavy quark mass from Goldstone boson loop.}
 \label{fig1}
\end{center}
\end{figure}

Such a gap equation was constructed recently
from a different, and in our view more \emph{ad hoc},
theoretical argument.
In Ref.~\cite{HX11}, an elementary Higgs doublet is assumed
together with a 4th generation.
Motivated by their earlier study~\cite{HX09}, where
some UV fixed point (UVFP) behavior was conjectured,
these authors pursued dynamical EWSB via a Schwinger-Dyson equation
that is rather similar to our Fig.~3.
However, perhaps in anticipation of the UVFP that might
develop at high energy~\cite{HX09}, they put in by hand
a \emph{massless} Higgs doublet, hence a scale invariant theory to boot.
It is the \emph{massless} Higgs doublet that runs in the loop,
replacing our Goldstone boson $G$.
The massless nature of the Higgs doublet appears \emph{ad hoc},
and the paper defers the discussion of the physical
Higgs spectrum for a future work.

In contrast, our Goldstone boson $G$, identified as the
collapsed $\pi_1$ state as it turns tachyonic,
is strictly massless in the broken phase.
In the gap equation of Fig.~3, we speculate that the
loop momentum should be cut off around $2m_Q$, rather than
some ``cut-off" scale $\Lambda > 2m_Q$.
In so doing, we bypass all issues of triviality that
arise from having $\Lambda$ approaching $2m_Q$.
What happens at scales above $2m_Q$ is to be studied
by experiment.

Here we remark that the first, elementary Higgs
of Ref.~\cite{HX11}, $\pi$ and $\sigma$,
are our bound state Goldstone bosons, and indeed
we should have a $\sigma$-like massive broad bound state
that could mimic the heavy Higgs boson.
Their second Higgs doublet, in the form of $t'$ and $b'$
bound states, would be excitations above the $\pi_1$
and $\sigma_1$ for us, likely rather broad.
We think that their claimed third doublet,
that of bound $\tau^\prime$ and $\nu_{\tau}^\prime$,
may not be bound at all, as their Yukawa couplings
may not be large enough.

The gap equation illustrated in Fig.~3 actually links to
a vast literature on strongly coupled, scale-invariant QED.
It is known that such a theory could have spontaneous
chiral symmetry breaking when couplings are strong enough.
Detailed numerical study would be deferred to a
future report, but Ref.~\cite{HX11} did provide some
numerical study of their gap equation, which,
up to some numerical factor, should be similar to ours.
So let us offer some remarks before concluding.

The integral gap equation can be cast into
differential form~\cite{FK76,BLL86} plus boundary conditions.
A critical value of $\lambda_Q \sim \sqrt{2}\pi$
seems to be required, which links to $m_Q \sim 770$ GeV.
This is above the current LHC bounds~\cite{CMS4th},
but not so far away!
The $\bar QQ$ condensate, the vacuum expectation value $v$
(the ``pion decay constant") can all be in principle
computed~\cite{PS79}.
The 770 GeV value is likely a lower bound on $m_Q$ for EWSB.
Furthermore, lacking the attraction of the scalar Higgs boson
as compared to Ref.~\cite{HX11},
we suspect that the effective Yukawa coupling, hence
critical $m_Q$, is likely larger.

Although our line of thought may seem constructed,
we have developed a self-consistent picture where
EWSB from large Yukawa coupling may be realized
with some confidence
 --- all without assuming an elementary Higgs boson.
We have not yet really touched on the $\sigma_1$ meson,
which would be the heavy Higgs boson.
However, our scenario is to have Goldstone bosons
as strongly (and tightly) bound ``Cooper pairs"
of very heavy quarks, which may please Nambu.

\section{Discussion and Conclusion}

We have adhered closely to an empirical approach
and collected several observations on the electroweak
sector and the possible mechanism of its breaking.
We showed that Yukawa couplings exist, as the couplings of
electroweak Goldstone bosons, or longitudinal vector bosons,
to fermions, which were deduced from purely left-handed
gauge couplings, utilizing the fact that both the vector
gauge bosons and the charged fermions are all massive.
This was done without any reference to the existence of
an elementary Higgs doublet.
With the backdrop that both the physical ``Higgs" particle
and the possible existence of an extra quark generation
seem to involve strong couplings, we discussed
Yukawa-bound $\bar QQ$ mesons.
Taking cue from the possible collapse of such states
at \emph{large} Yukawa coupling, we {postulate} that
the collapsed MAC state, the $\pi_1$ isotriplet, color-singlet
meson is the Goldstone boson $G$ (or $W_L$) itself.
This $\pi_1$ Goldstone boson is a rather tightly
bound (very small radius) state.
From this picture, we constructed a simple gap equation
that is quite similar to strongly coupled, scale-invariant
QED, which is known to exhibit dynamical chiral symmetry
breaking at large coupling. A rough estimate of the
minimal critical $m_Q \sim 800$ GeV was suggested.

In essence, we suggest a ``bootstrap" picture
where both the Goldstone boson,
and the heavy quark mass, are generated by
a strongly coupled gap equation.
The discussion has been heuristic, and
the dynamical EWSB is rooted in the existence
of a strong Yukawa coupling, without
offering any theory for this coupling,
except that the existence of Yukawa couplings
is based on experiment.
Unlike the Nambu--Jona-Lasinio (NJL) model~\cite{NJL},
there is no assumption of an effective 4-quark operator.
Instead, one utilizes nothing but the Yukawa coupling itself.
In the NJL model, the Goldstone boson is a ladder sum
of quark-antiquarks interacting through the effective
4-quark operator, but in our case, the interaction
is through the Goldstone boson itself, hence
closer to a bootstrap model.
It would be interesting to explore further the
similarities and interconnections of our picture
with the NJL model approach~\cite{BHL}.

Chiral symmetry breaking in hadron physics is
realized by QCD. There are striking differences
between QCD (hence technicolor-like models of EWSB)
and strong Yukawa induced dynamical EWSB:
the heavy quark $Q$ is not confined.
Also, in the QCD picture, the physical pion is
still a stringy state.
For our $\pi_1$, we do not know how it would be
finally realized in an ultimate theory
(that would explain the origin of Yukawa couplings).

There is an experimental perspective on our gap equation.
The longitudinal component of the electroweak vector
boson, $V_L$, is the Goldstone boson $G$.
The striking success of the Standard Model against
all other alternative, New Physics theories, after collecting
two times 5 fb$^{-1}$ data at the 7 TeV LHC, makes clear that
there may be no other object around or not too far above the weak scale.
Thus, in the gap equation of Fig.~3, the Goldstone boson
loop indeed seems to sum up all dominant effects, if
there exist some yet unseen very heavy quark $Q$.
Thus, this view of dynamical EWSB is consistent with
2011 LHC data, and might be testable in the not so distant future.

In conclusion,
with heuristic arguments of the physical nature
of Yukawa couplings, but without touching on
an elementary Higgs sector, by introducing
large Yukawa couplings through an extra generation
of chiral quarks, we have illustrated how the
electroweak symmetry may be dynamically broken,
where the gap equation implies $\bar QQ$
condensation as the origin of EWSB,
and a strongly coupled ``Higgs" sector would
emerge, with a corresponding spectrum of
heavy $\bar QQ$ states.
This view may become relevant once the
current hint of a 125 GeV light Higgs boson
is disproved by experiment, which can happen
by the end of 2012.

\vskip0.3cm
\noindent {\bf Acknowledgement}
 I thank
 J. Alwall, J.-W. Chen, K.-F. Chen, H.-C. Cheng, T.-W. Chiu,
 T. Enkhbat, K.~Jensen, H.~Kohyama, Y.~Kikukawa,, C.-J.D.~Lin,
 F.J.~Llanes~Estrada, Y. Mimura and H. Yokoya for discussions,
 and special thanks to T.~Kugo for reading the manuscript.
 Any shortcoming is of course mine.
 This research is supported by the Academic Summit grant
 NSC 100-2745-M-002-002-ASP of the National Science Council of Taiwan.


\begin{thebibliography}{99}
%
%
\bibitem{Peskin}
  Summary talk by M.E. Peskin at Lepton Photon Symposium,
  August 2011, Mumbai, India [arXiv:1110.3805 [hep-ph]].
%
\bibitem{EPSHiggs}
  Plenary talks by G. Tonneli, D. Charlton and B. Murray
  at Europhysics Conference on High-Energy Physics, July 2011,
  Grenoble, France.
%
\bibitem{LPHiggs}
  Plenary talks by A. Nisati and V. Sharma at
  Lepton Photon Symposium, August 2011, Mumbai, India.
%
\bibitem{Dec13Higgs}
  Special seminars delivered by F.~Gianotti and G.~Tonneli on
  December 13, 2011, CERN, Geneva, Switzerland.
%
\bibitem{LQT77}
  B.W.~Lee, C.~Quigg and H.B.~Thacker,
  Phys.\ Rev.\ Lett.\  {\bf 38}, 883 (1977);
  Phys.\ Rev.\  D {\bf 16}, 1519 (1977).
%
\bibitem{CFH78}
  M.S.~Chanowitz, M.A.~Furman and I.~Hinchliffe,
  Phys.\ Lett.\  B {\bf 78}, 285 (1978);
  Nucl.\ Phys.\  B {\bf 153}, 402 (1979).
%
\bibitem{CMS4th}
  Public Analysis Summaries (PAS) CMS-EXO-11-036, CMS-EXO-11-050,
  CMS-EXO-11-051 and CMS-EXO-11-054 of the CMS Collaboration.
%
\bibitem{Higgs4}
  Talks by G. Tonneli in Ref.~\cite{EPSHiggs},
  and A. Nisati in Ref.~\cite{LPHiggs}.
%
\bibitem{Denner}
  In fact, cross section enhancement when both the Higgs
  and $Q$ are very heavy demands closer scrutiny.
  For a recent discussion, see
  A.~Denner {\it et al.},
  arXiv:1111.6395 [hep-ph].
%
\bibitem{Holdom06}
  See \emph{e.g.} B.~Holdom,
  JHEP {\bf 0608}, 076 (2006), and references therein.
%
\bibitem{PDG}
  K. Nakamura \textit{et al.} [Particle Data Group], J. Phys. G \textbf{37}, 075021 (2010).
%
\bibitem{tHV72}
  G.~'t Hooft and M.J.G.~Veltman,
  Nucl.\ Phys.\  B {\bf 44}, 189 (1972).
%
\bibitem{'tHooft71}
  G.~'t Hooft,
  Nucl.\ Phys.\  B {\bf 35}, 167 (1971).
  The author attributes the demonstration for the global gauge
  symmetry case to work done by B.W. Lee, K. Symanzik, and J.-L. Gervais.
%
\bibitem{HouICHEP10}
  A brief account was given earlier in
  G.W.-S.~Hou,
  PoS {\bf ICHEP2010}, 244 (2010)
  [arXiv:1101.2158 [hep-ph]].
%
\bibitem{KugoComment}
  Strictly speaking, only the on-shell $S$-matrix is finite
  in $U$-gauge. We thank T. Kugo for the comment,
  and take it as a limiting case of the
  $R_\xi$ gauge for sake of illustration.
%
\bibitem{KM}
 M. Kobayashi and T. Maskawa,
  Prog.\ Theor.\ Phys.\ {\bf 49}, 652 (1973).
%
\bibitem{4S4G09}
  For a recent brief review, see
  B.~Holdom, W.-S.~Hou, T.~Hurth, M.L.~Mangano, S.~Sultansoy and G.~\"{U}nel,
  PMC Phys.\  A {\bf 3}, 4 (2009).
%
\bibitem{KPST07}
 G.D. Kribs, T. Plehn, M. Spannowsky, and T.M.P. Tait,
  Phys.\ Rev.\ D {\bf 76}, 075016 (2007).
%
\bibitem{Wise11}
  K.~Ishiwata and M.B.~Wise,
  Phys.\ Rev.\  D {\bf 83}, 074015 (2011).
%
\bibitem{Jain}
  P.~Jain, D.W.~McKay, A.J.~Sommerer, J.R.~Spence, J.P.~Vary and B.-L.~Young,
  Phys.\ Rev.\  D {\bf 46}, 4029 (1992);
  \emph{ibid.}\  D {\bf 49}, 2514 (1994).
%
\bibitem{EHY11}
  T.~Enkhbat, W.-S.~Hou and H.~Yokoya,
  Phys. Rev. D \textbf{84}, 094013 (2011).
%
\bibitem{Kugo78}
  T.~Kugo,
  Phys.\ Lett.\  B {\bf 76}, 625 (1978).
%
\bibitem{omega1}
  There is another aspect on the self-consistency of this subtraction/Postulate. 
  If one did not remove the $\pi_1$ equation by invoking this subtraction, 
  the leading collapse state of strong Yukawa coupling would be the $\omega_1$. 
  Condensation in this channel would break Lorentz invariance.
%
\bibitem{HX11}
  P.Q.~Hung and C.~Xiong,
  Nucl.\ Phys.\  B {\bf 848}, 288 (2011).
%
\bibitem{HX09}
  P.Q.~Hung and C.~Xiong,
  Phys.\ Lett.\  B {\bf 694}, 430 (2011);
  Nucl.\ Phys.\  B {\bf 847}, 160 (2011).
%
\bibitem{FK76}
  R.~Fukuda and T.~Kugo,
  Nucl.\ Phys.\  B {\bf 117}, 250 (1976).
%
\bibitem{BLL86}
  W.A.~Bardeen, C.N.~Leung and S.T.~Love,
  Phys.\ Rev.\ Lett.\  {\bf 56}, 1230 (1986);
  C.N.~Leung, S.T.~Love and W.A.~Bardeen,
  Nucl.\ Phys.\  B {\bf 273}, 649 (1986).
%
\bibitem{PS79}
  H.~Pagels and S.~Stokar,
  Phys.\ Rev.\  D {\bf 20}, 2947 (1979).
%
%
\bibitem{NJL}
  Y.~Nambu and G.~Jona-Lasinio,
  Phys.\ Rev.\  {\bf 122}, 345 (1961);
  {\it ibid}\  {\bf 124}, 246 (1961).
%
\bibitem{BHL}
  W.A.~Bardeen, C.T.~Hill and M.~Lindner,
  Phys.\ Rev.\  D {\bf 41}, 1647 (1990).
%
\end{thebibliography}
\end{document}